\documentclass[]{raa}            
\usepackage{graphicx,times}
\usepackage{natbib}
\usepackage{epstopdf}

\providecommand{\bm}[1]{\mbox{\boldmath$#1$\unboldmath}}

\begin{document}

   \title{Photometric Observations of Three High Mass X-Ray Binaries and a Search for Variations Induced by Orbital
Motion}

 \volnopage{ {\bf 20xx} Vol.\ {\bf 9} No. {\bf XX}, 000--000}
   \setcounter{page}{1}

   \author{Gordon E. Sarty
      \inst{1}
   \and Bogumil Pilecki
      \inst{2,3}
   \and Daniel E. Reichart
      \inst{4}
   \and Kevin M. Ivarsen
      \inst{4}
   \and Joshua B. Haislip
      \inst{4}
   \and Melissa C. Nysewander
      \inst{4} 
   \and Aaron P. LaCluyze
      \inst{4}
   \and Helen M. Johnston
      \inst{5}
   \and Robert R. Shobbrook
      \inst{5} 
   \and L\'{a}szl\'{o} L. Kiss
      \inst{6}
   \and Kinwah Wu
     \inst{5,7}
   }

   \institute{Departments of Psychology and Physics, University of Saskatchewan, 9 Campus Drive, Saskatoon, Saskatchewan S7N 5A5, Canada; {\it gordon.sarty@usask.ca}\\
        \and
             Warsaw University Observatory, Al. Ujazdowskie 4, 00-478 Warsaw, Poland
        \and
             Universidad de Concepcion, Departamento de Astronom\"{i}a, Casilla 160-C,
Concepcion, Chile
        \and
             Department of Physics and Astronomy, University of North Carolina -- Chapel Hill, CB 3255, Phillips Hall, Chapel Hill, NC 27599-3255, USA
        \and
             School of Physics A28, University of Sydney, New South Wales 2006, Australia
        \and
             Konkoly Observatory, H-1525 Budapest, P.O. Box 67, Hungary
        \and
             Mullard Space Science Laboratory, University College London, 
Holmbury St.~Mary, Dorking, Surrey RH5 6NT, United Kingdom \\
\vs \no
   {\small Received [year] [month] [day]; accepted [year] [month] [day] }
}

\abstract{We searched for long period variation in $V$-band, $I_{C}$-band and {\em RXTE} X-ray light curves of the High Mass X-ray Binaries (HMXBs) LS 1698 / RX J1037.5$-$5647, HD 110432 / 1H 1249$-$637 and HD 161103 / RX J1744.7$-$2713 in an attempt to discover orbitally induced variation. Data were obtained primarily from the ASAS database and were supplemented by shorter term observations made with the 24- and 40-inch ANU telescopes and one of the robotic PROMPT telescopes. Fourier periodograms suggested the existence of long period variation in the $V$-band light curves of all three HMXBs, however folding the data at those periods did not reveal convincing periodic variation. At this point we cannot rule out the existence of long term $V$-band variation for these three sources and hints of longer term variation may be seen in the higher precision PROMPT data. Long term $V$-band observations, on the order of several years, taken at a frequency of at least once per week and with a precision of 0.01 mag, therefore still have a chance of revealing long term variation in these three HMXBs.
}

   \authorrunning{G.E. Sarty et al. }            
   \titlerunning{Three Long Period X-ray Binaries}  
   \maketitle


%
%
\section{Introduction}\label{sec1}

High Mass X-ray Binaries (HMXBs) consist of a massive star orbited by a compact object like a neutron star, a black hole or, as is proposed for $\gamma$ Cas analogues, a white dwarf. A major class of HMXBs, the BeXs, have a Be star as the primary massive star. The orbital periods of many BeXs are known from their periodic Type I X-ray outbursts. Such outbursts are believed to be caused by the passage of a neutron star through the Be star's equatorial decretion disk when increased accretion onto the neutron star occurs \citep{okazaki2001}. These Type I X-ray outbursts can also be accompanied by optical outbursts which can confirm the orbital period \citep{coe2004}. However, many BeXs including the three of interest here, are persistent X-ray sources, indicating a more uniform accretion rate and circular orbits with longer periods. The lack of regular X-ray outbursts means that the orbital periods for many of the persistent BeXs are unknown. In spite of the expected lack of corresponding optical outbursts, optical searches for orbital periodicity can be successful \citep{coe2005,Schmidtke2006}.

Here we report on the search for long periods in the optical photometry of three persistent BeXs: LS 1698, the optical counterpart to RX J1037.5$-$5647; HD 110432, the optical counterpart to 1H 1249$-$637 and; HD 161103, the optical counterpart to RX J1744.7$-$2713. The source LS 1698 has been argued to be an analogue of X Per \citep{reig1999} while HD 110432 and HD 161103 have been argued to be analogues of $\gamma$ Cas. X Per is a BeX with a neutron star and a long orbital period of 250.3 d \citep{Delgado2001}. The star $\gamma$ Cas has been proposed to contain a white dwarf as its compact object, although single star models for $\gamma$ Cas exist \citep{oliveria2006,oliveria2007}.

The main component of our data came from the All Sky Automated Survey (ASAS, \citet{pojmanski1998}) supplemented with confirming data obtained with the 24 and 40 inch telescopes of the Australian National University (ANU) and with one of the 16 inch telescopes of the Panchromatic Robotic Optical Monitoring and Polarimetry Telescopes (PROMPT, \citet{reichart2005}) array.

This paper is organized as follows. In Section \ref{sec2} we present the photometric observations, including {\em RXTE} X-ray photometry for LS 1698. Section \ref{sec3} gives the results of formal searches for periods in the data using Fourier techniques. In Section \ref{sec4} we give a brief review of what is currently known about the three HMXBs to motivate further investigation of these sources. Our conclusions are summarized in Section \ref{sec5}.

\section{Observations}\label{sec2}

\begin{figure}
\begin{center}
\includegraphics[angle=0,scale=0.42]{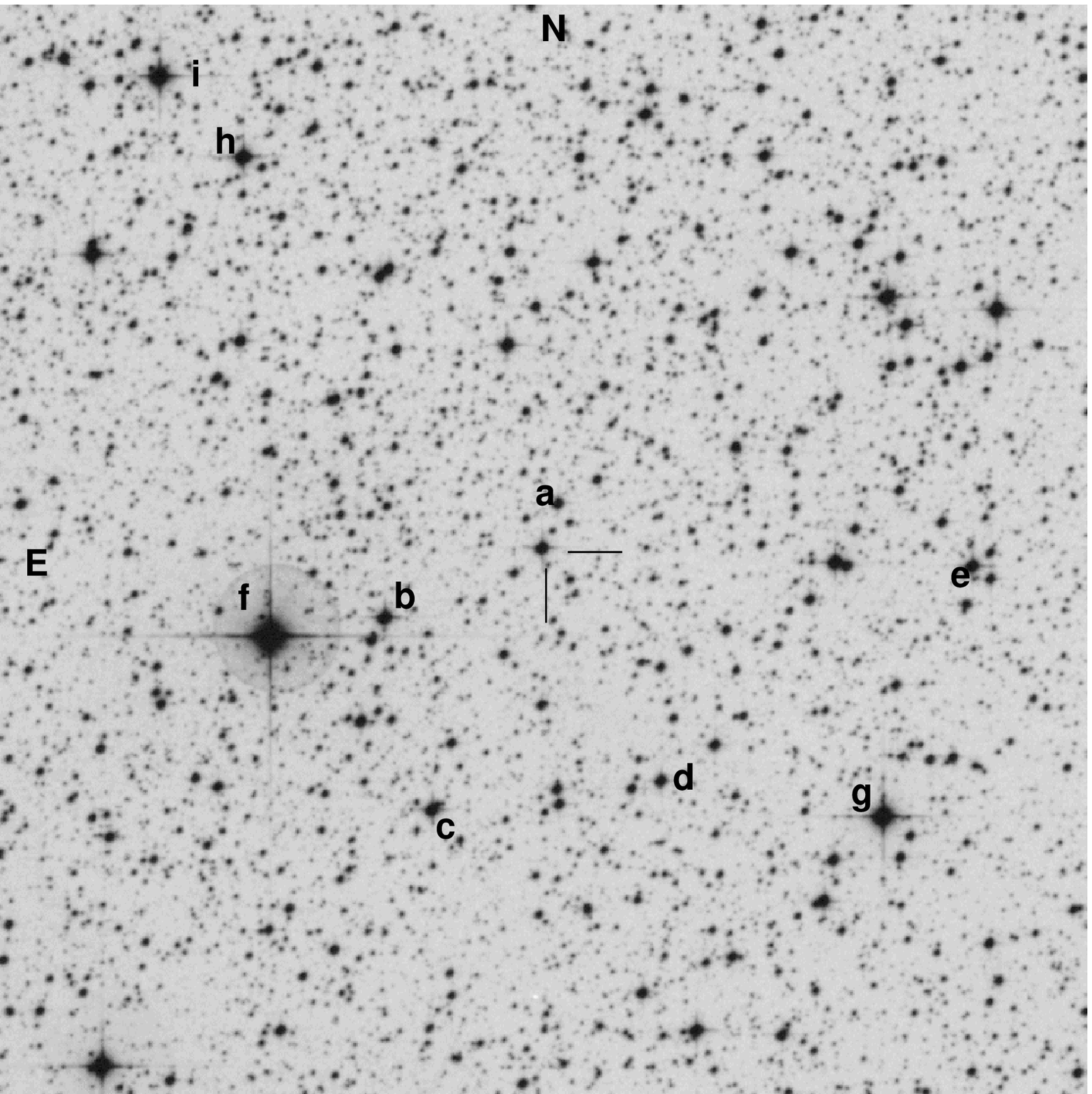}

\begin{center}
\scriptsize
\begin{tabular}{ccccccc}
\hline
\multicolumn{7}{c}{Comparison/Check stars} \\
\hline
Star & RA          & Dec          &  $B$    & $V$   & $R$   &  $I_{C}$  \\
a   & 10 37 33.4   &  -56 47 19.4 & 12.82	& 12.45 & 12.12 &  11.71  \\
c	& 10 37	47.17  &  -56 51 29.2 &	11.62	& 11.17 & 10.81 & 10.41 \\
d	& 10 37	23.9   &  -56 51 11.4 &	12.52	& 12.05 & 11.67 & 11.28 \\
e	& 10 36	52.05  &  -56 48 20.9 &	11.54	& 11.18 & 10.83 & 10.41 \\
\hline
\end{tabular}
\normalsize
\end{center}
\end{center}
\caption{Comparison/check stars in the field of LS 1698 as determined from data obtained with the ANU 24- and 40-in telescopes at Siding Spring Observatory. Errors are on the order of 0.01 to 0.05 mag. The target star is marked with two lines in the centre. Positions and magnitudes are listed for those stars whose magnitudes and colours make them most useful for differential photometry with the target. North is up, east to the left. \label{stdstars69}}
\end{figure}

\begin{figure}
\begin{center}
\includegraphics[angle=0,scale=0.42]{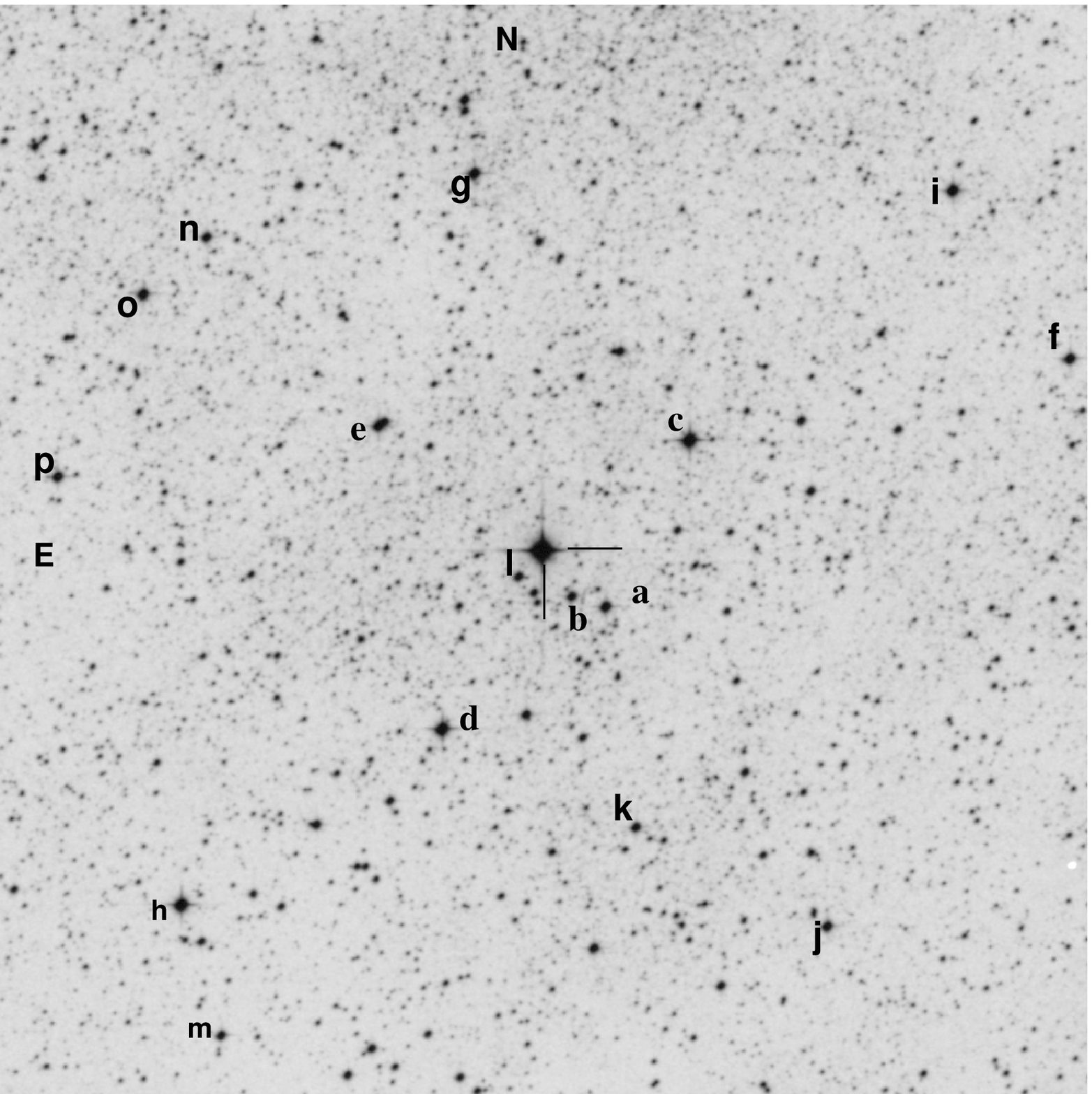}

\begin{center}
\scriptsize
\begin{tabular}{ccccccc}
\hline
\multicolumn{7}{c}{Comparison/Check stars} \\
\hline
Star & RA      & Dec         &  $B$   & $V$    &  $R$   & $I_{C}$  \\
a & 17 44 41.9 & -27 14 31.5 & 12.881 & 12.082 & 11.587 & 11.053 \\
b & 17 44 44.0 & -27 14 22.5 & 13.056 & 12.590 & 12.311 & 12.002 \\
d & 17 44 52.1 & -27 16 10.5 & 11.298 & 10.731 & 10.427 & 10.081 \\
f & 17 44 12.9 & -27 11 11.5 & 12.525 & 11.936 & 11.602 & 11.228 \\
j & 17 44 28.4 & -27 18 57.4 & 12.849 & 12.222 & 11.833 & 11.392 \\
k & 17 44 40.2 & -27 17 33.7 & 13.281 & 12.648 & 12.216 & 11.817 \\
l & 17 44 47.3 & -27 14 05.5 & 13.265 & 12.716 & 12.352 & 11.983 \\
\hline
\end{tabular}
\normalsize
\end{center}
\end{center}
\caption{Comparison/check stars in the field of HD 161103 as determined from data obtained at the Sonoita Research Observatory (SRO) by Arne Henden of the AAVSO. Errors are on the order of 0.001 to 0.005 mag. The target star is marked with two lines in the centre. Positions and magnitudes are listed for those stars whose magnitudes and colours make them most useful for differential photometry with the target. North is up, east to the left. \label{stdstars95}}
\end{figure}

\begin{table}
\caption{Observation Log. \label{tabobs}}
\begin{center}
\begin{tabular}{cccc}
\hline
Source   &  Dates$^{1}$           &  Filters         &  Telescope \\
\hline
LS 1698 & 21/12/2000 to 3/12/2009 & $V$,$I_{C}$          &  ASAS  \\ 
        & 24/8/2005 to 5/9/2005   & $B$,$V$,$R$,$I_{C}$  & ANU 40-in \\
        & 24/3/2006 to 1/4/2006   & $B$,$V$,$R$,$I_{C}$  & ANU 24-in \\
        & 17/7/2007 to 28/7/2007  & $B$,$V$,$R$,$I_{C}$  & ANU 40-in \\
        & 8/12/2008 to 12/2/2009  & $V$,$I_{C}$          & PROMPT-5 \\
HD 110432 & 29/11/2000 to 29/7/2009 & $V$,$I_{C}$          &  ASAS \\ 
HD 161103 & 31/1/2001 to 2/7/2009 & $V$,$I_{C}$          &  ASAS \\ 
          & 4/8/2002 to 14/10/2002 & $B$,$V$,$I_{C}$  & ANU 24-in \\ 
          & 29/8/2005 to 5/9/2005 & $B$,$V$,$R$,$I_{C}$  & ANU 40-in \\
          & 23/3/2006 to 1/4/2006 & $B$,$V$,$R$,$I_{C}$  & ANU 24-in \\
          & 19/7/2006 to 19/7/2006 & $B$,$V$,$R$,$I_{C}$  & ANU 40-in \\
\hline
\multicolumn{4}{l}{$^{1}$ day/month/year}
\end{tabular}
\end{center}
\end{table}

A total of four small telescopes/telescope systems were used to make optical observations of LS 1698, HD 110432 and HD 161103. The observation log is given in Table \ref{tabobs}. 

The bulk of the data were obtained with the ASAS-3 telescope system. ASAS-3 consists of two wide field 87-mm diameter f2.8 Minolta telephoto lenses, one equipped with a Johnson $V$ filter, the other with the Johnson-Cousins $I_{C}$ filter. The individual telescopes use Apogee AP-10 2048$\times$2048 pixel CCD cameras as detectors.  ASAS-3 is an automated instrument and produces reduced photometry automatically. As part of the reduction process, the software rates the quality of the data. Only observations rated as `grade A' were used. The ASAS instrument is located at the Las Campanas Observatory in Chile.

Two telescopes owned by the ANU and located at the Siding Spring Observatory in Australia were used. These telescopes had apertures of 24 and 40 inches. 

\begin{figure*}[h]
\begin{center}
\includegraphics[angle=0,scale=1]{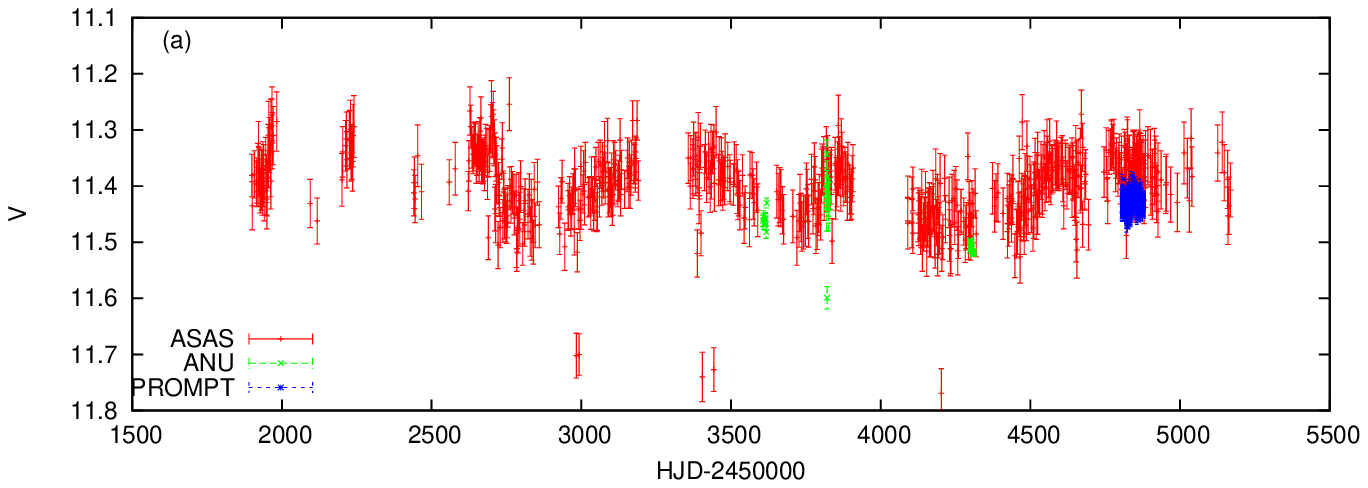}

\includegraphics[angle=0,scale=1]{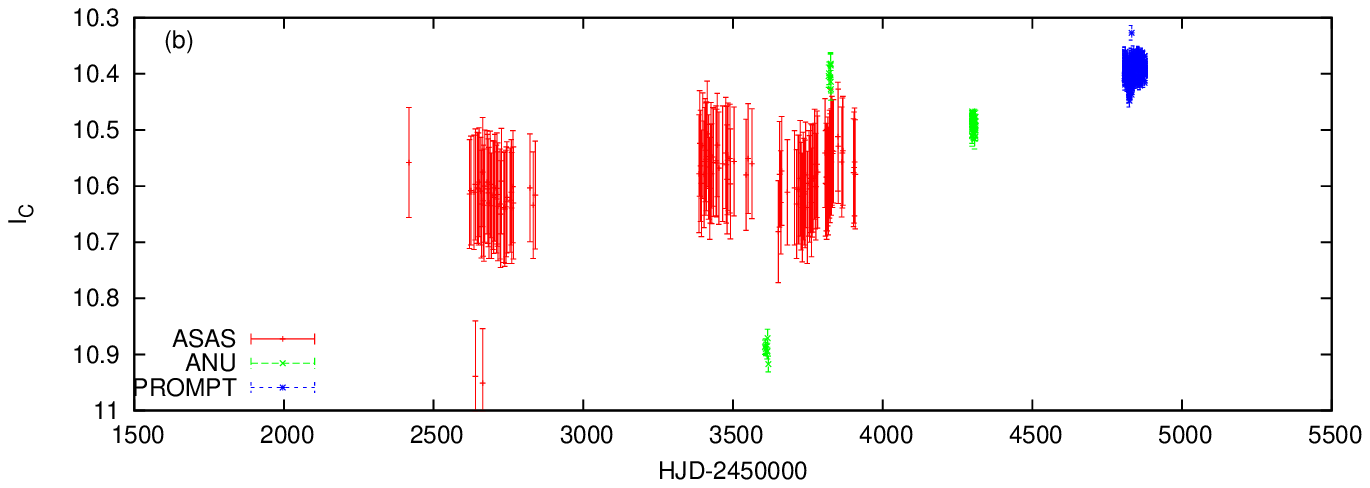}

\end{center}
\caption{(a) LS 1698 $V$ band data. (b) LS 1698 $I_{C}$ band data. A few outlier ASAS data points below $V = 11.8$ and $I_{C} = 11.0$ have been cut-off in the plots.\label{LPH069Vfig}}
\end{figure*}

\begin{figure*}[h]
\begin{center}
\includegraphics[angle=0,scale=1]{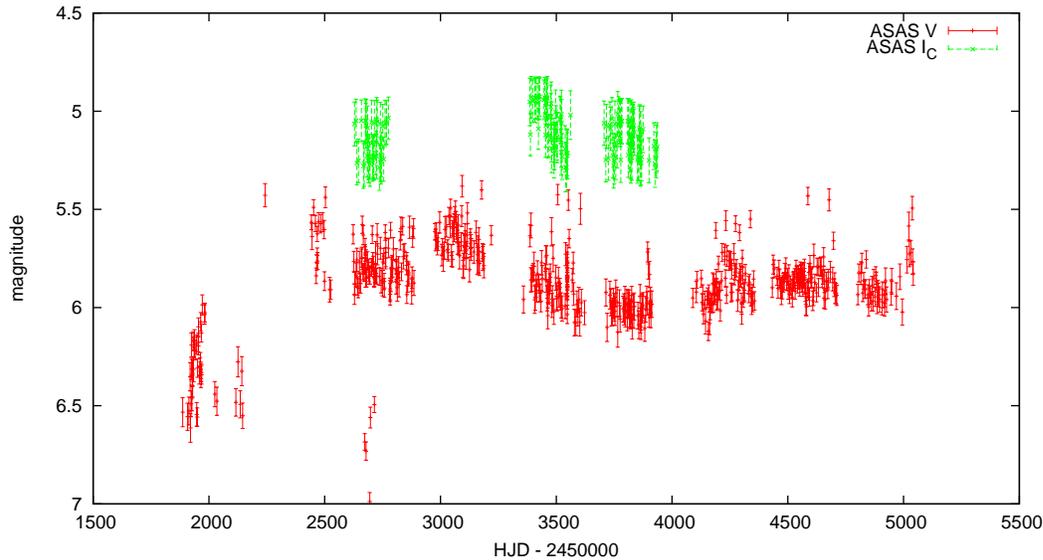}

\end{center}
\caption{HD 110432 $V$ and $I_{C}$ band data. A few $V$-band outliers below 7th magnitude have been cut-off in the plot. The data before HJD 2452300 were exposed with a different protocol than later data and non-linear CCD response for this bright source may be responsible for the apparently discrepant data points at the beginning of the light curve. \label{LPH079Vfig}}
\end{figure*}

\begin{figure*}[h]
\begin{center}
\includegraphics[angle=0,scale=1]{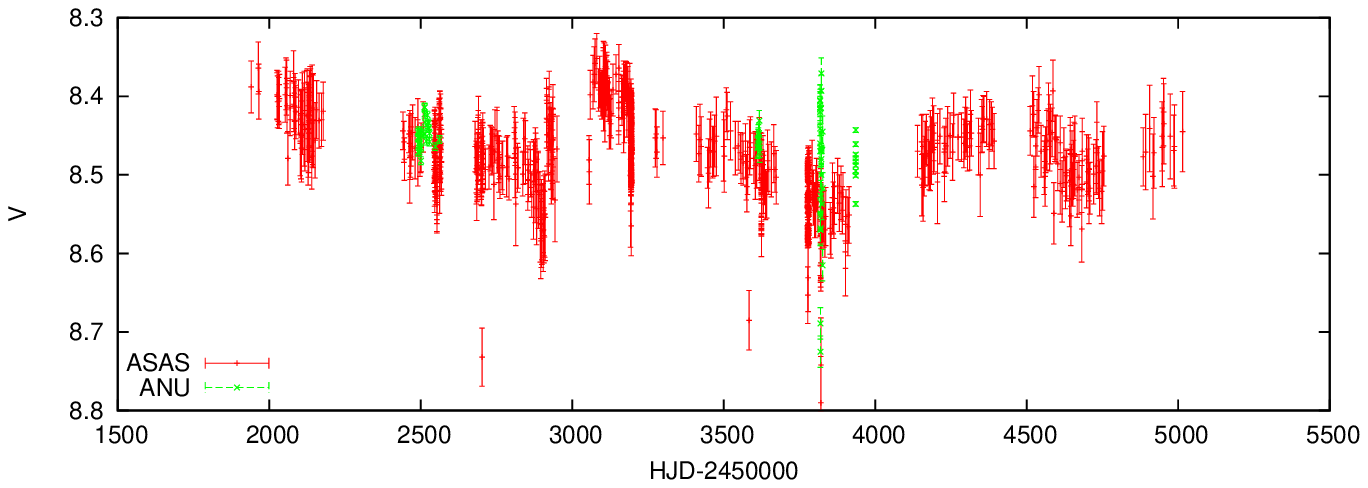}

\includegraphics[angle=0,scale=1]{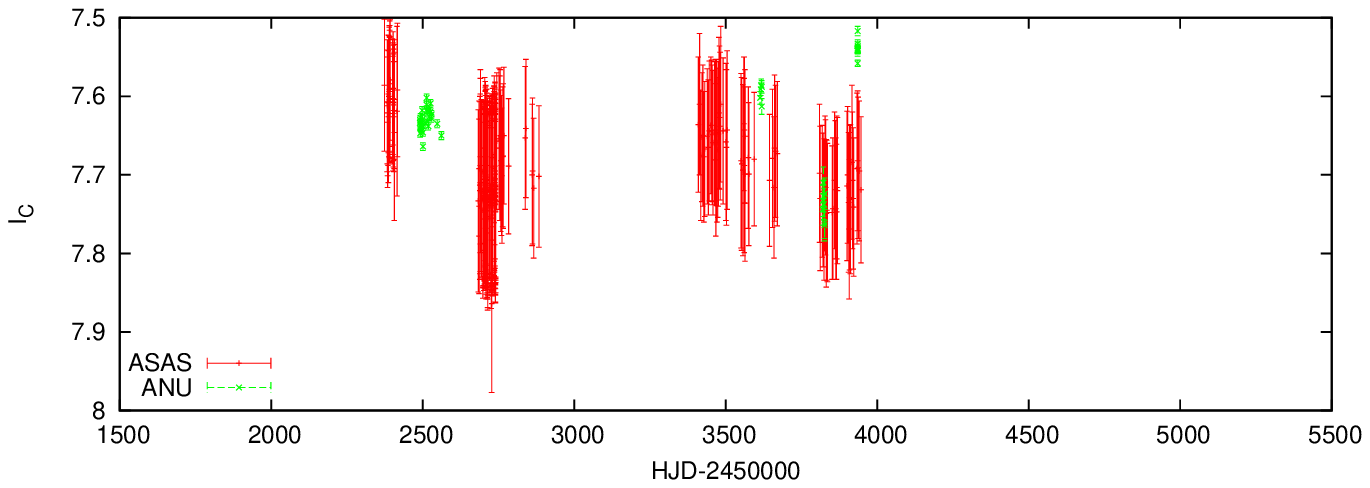}
\end{center}
\caption{(a) HD 161103 $V$ band data. (b) HD 161103 $I_{C}$ band data. A few ASAS outliers below $V = 8.8$ have been cut-off in the plot. \label{LPH095Vfig}}
\end{figure*}

For the first observing run with the ANU 24-in telescope, from 4/8/2002 to 14/10/2002, for HD 161103 a pulse-counting GaAs photo-multiplier tube photometer was used with a filter wheel controlled by a computer that also recorded photon counts in the $B$, $V$ and $I_{C}$ bands \citep{bessel1990}. Transformations to the standard Cape $UBVRI$ system were made on one night from 15 stars in the Regions E5, E7 and E9, using the values of \citet{menzies1989}. On each night, the $V$, $(B-V)$ and $(V-I)$ magnitudes of LS 1698 were refered to the two standards E6/04 and E6/66, which are approximately 18$^{\circ}$ from the source. The internal photometric errors of the observations were of the order 0.002 to 0.004 mag. Referral to the fairly distant E6 Region stars adds an uncertainty of similar amount on average, depending on the air-mass and the quality of the night, due to the uncertainty in the differential extinction corrections. The final differential photometric errors were therefore in the range 0.005 to 0.007 mag. 

Subsequent observing runs with the ANU 24-in telescope, for LS 1698 and HD 161103, employed an imaging camera with a SiTE 2048$\times$2048 pixel LN$_{2}$ cooled CCD and a manually turned filter wheel with $B$, $V$, $R$ and $I_{C}$ filters. The camera was controlled with the {\small CICADA} software \citep{young1999}. Standard stars in the Landolt SA104 field \citep{landolt1992} were observed for the purpose of defining comparison stars in the field of LS 1698. From the Landolt SA104 stars, extinction and filter band transformation coefficients were computed and combined with similar data from the ANU 40-in telescope (see below) to define the magnitudes of standard stars in the LS 1698 field (see Fig.~\ref{stdstars69}). Standard star magnitudes in the field of HD 161103 were determined from data obtained at the Sonoita Research Observatory (SRO) and were provided by Arne Henden of the American Association of Variable Star Observers (AAVSO) (see Fig.~\ref{stdstars95}). The errors on the magnitudes of the standard stars in the LS 1698 field were in the range 0.01 to 0.05 mag. The errors on the target star magnitudes, as determined by the inhomogeneous differential photometry technique \citep{honeycutt1992}, were generally 0.005 mag (see Figs.~\ref{LPH069Vfig} to \ref{LPH095Vfig}). Inhomogeneous differential photometry determines the differential magnitudes of all analysed stars at once, using a least squares approach, and references the differences relative to one selected star. For LS 1698, the reference star was star a as labelled in Fig.~\ref{stdstars69}. For HD 161103, the reference star was star c as labelled in Fig.~\ref{stdstars95}. The magnitudes of the reference stars were therefore added to the differential magnitudes as determined by the inhomogeneous photometry technique. 

The 40-in telescope was equipped with the Wide Field Imager (WFI, \citet{baade1998}) and a filter wheel with $B$,$V$,$R$ and $I_{C}$ filters. The LN$_{2}$ cooled camera and filter wheel were controlled with the {\small CICADA} software. Similar to the observing procedure described for the 24-in ANU telescope, standard stars in the Landolt SA104 and SA113 fields were observed for the purpose of defining comparison stars in the source fields. As with the description given for the 24-in telescope above, extinction and filter band transformation coefficients were computed, comparison star magnitudes determined, and source magnitudes determined using the inhomogeneous photometry approach with similar errors. The differentially determined magnitudes of the target stars, for both the 24-in and 40-in telescope data, were not transformed to a standard filter band system since the precision of the transformation coefficients was deemed adequate only for determining comparison star magnitudes and not for use in differential photometry. As a check on the comparison star magnitude determinations, a comparison with magnitudes published in the Simbad database was made where possible, with excellent agreement between the determined and published magnitudes. 

The fourth telescope used was the PROMPT-5 telescope located at the Las Campanas Observatory in Chile. The PROMPT telescope system is a robotic system consisting of five 16-in Richey-Chretien telescopes built by RC Optical Systems. Each telescope is equipped with a fast readout ($<1$ s) Apogee Alta U47+ 1024$\times$1024 CCD camera. The PROMPT system is under the control of Skynet, a prioritized queue scheduling and control program running on a computer at UNC-Chapel Hill's Morehead Observatory. Skynet interacts with MySQL databases and commands dumb-by-design `Terminator' programs at each telescope. Images are automatically transferred back to a 1.1 terrabyte RAID 5 array at Morehead. 

Calibration of the CCD data for the ANU telescope and PROMPT data, involving bias subtraction, flat fielding from sky flats taken in morning and evening twilight and, for the PROMPT data, dark frame subtraction, were performed with the {\small IRAF}\footnote{{\small IRAF} is the Image Reduction and Analysis Facility, a general purpose software system for the reduction and analysis of astronomical data. {\small IRAF} is written and supported by the {\small IRAF} programming group at the National Optical Astronomy Observatories (NOAO) in Tucson, Arizona. NOAO is operated by the  Association of Universities for Research in Astronomy (AURA), Inc. under cooperative agreement with the National Science Foundation}  software and the {\bf ccdproc} task. Instrumental magnitudes for the ANU telescope and PROMPT data were determined with the point-spread-function approach \citep{stetson1987} using the {\bf psf} and {\bf allsky} tasks of the {\small IRAF} software. The instrumental magnitudes were then subjected to the inhomogeneous differential photometry approach as previously described.

Data taken for the three sources in the $V$ and $I_{C}$ bands are shown in Figs.~\ref{LPH069Vfig} to \ref{LPH095Vfig}. Data from the $B$ and $R$ bands were not of sufficient temporal density to be useful for period determination; they are summarized in Table \ref{tabBR} for completeness sake.

\begin{figure}
\begin{center}
\includegraphics[angle=0,scale=1]{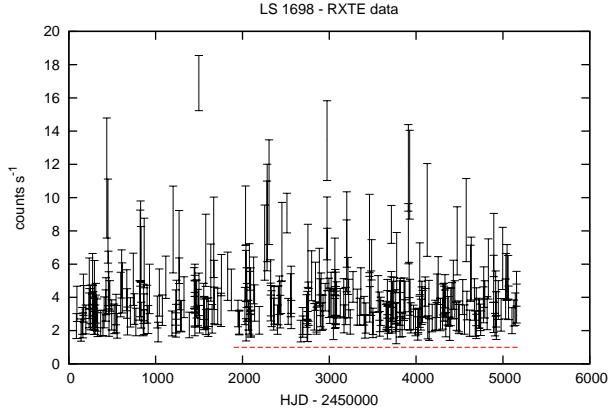}
\end{center}
\caption{2-10 keV {\em RXTE} light curve for LS 1698 / RX J1037.5$-$5647. The red line indicates the time-span of of the optical observations shown in Fig.~\ref{LPH069Vfig}. \label{RXTELPH069fig}}
\end{figure}

X-ray data for RX J1037.5$-$5647, for which LS 1698 is the optical counterpart, were available from the {\em Rossi X-Ray Timing Explorer} ({\em RXTE}) satellite. Definitive data (as opposed to quick-look data) from the All Sky Monitor (ASM) instrument were downloaded from the {\em RXTE} Guest Observer Facility for the purposes of finding correlations with the optical wavelength variations. The {\em RXTE} ASM is composed of three Scanning Shadow Cameras (SSCs) that perform sets of 90 s pointed observations (dwells) covering about 80 per cent of the sky every $\sim$90 min \citep{levine1996}. The data used were from the dwell by dwell compilation. Each raw data point represents the fitted source flux from one 90 s dwell. Data from all three SSCs were used and represent nominal 2-10 keV rates in ASM counts s$^{-1}$. Nominally, the Crab nebula flux is about 75 ASM counts s$^{-1}$ (when the source is at the centre of an SSC field of view and all eight anodes are operational). We retained only the `3$\sigma$' detections based on the reported variance for each data point. The ASM data for LS 1698 / RX J1037.5$-$5647 are shown in Fig.~\ref{RXTELPH069fig}.

\begin{table}
\caption{Summary of $B$ and $R$ band data from LS 1698 and HD 161103. \label{tabBR}}
\scriptsize
\begin{center}
\begin{tabular}{cccccc} 
\hline
Source  & band & Time span$^{1}$  & Mean  & Std.  & No.\\
        &        &                  &  mag  & Dev. & Obs.\\
\hline
LS 1698 & $B$    & 24/8/2005 to 5/9/2005 &  12.004 &  0.015 & 8 \\
        &        & 24/3/2006 to 1/4/2006 &  11.947 &  0.048 & 6 \\
        &        & 17/7/2007 to 28/7/2007 & 12.043 &  0.012 & 26 \\
        & $R$    & 24/8/2005 to 5/9/2005 &  10.983 &  0.010 & 8 \\
        &        & 24/3/2006 to 1/4/2006 &  10.976 &  0.053 & 10 \\
        &        & 17/7/2007 to 28/7/2007 & 11.037 &  0.009 & 29 \\
HD 161103 & $B$  & 4/8/2002 to 14/10/2002 &  8.891 &  0.017 & 29 \\
          &      & 29/8/2005 to 5/9/2005  & 8.898 &  0.135 & 10 \\
          &      & 23/3/2006 to 1/4/2006  & 9.223 &  0.027 & 3 \\
          &      & 19/7/2006 to 19/7/2006 & 8.904 &  0.006 & 11 \\
          & $R$  & 29/8/2005 to 5/9/2005 &  8.041  & 0.017 & 11 \\
          &      & 23/3/2006 to 1/4/2006  & 8.081  & 0.037 & 9  \\
          &      & 19/7/2006 to 19/7/2006 & 8.058  & 0.010 & 11  \\
\hline 
\multicolumn{4}{l}{$^{1}$ day/month/year}
\end{tabular}
\end{center}
\normalsize
\end{table}

\clearpage

\section{Results}\label{sec3}

Fourier analysis of the photometric data was performed using the software {\footnotesize PERIOD04} \citep{lenz2005}. Analyses were performed with either the data weighted by the inverse of the errors or with equally weighted data. Unless otherwise reported (e.g. for the {\em RXTE} data) analyses with data weighted by the inverse of the errors did not produce any sensible results. The largest peak was deemed significant if the signal to noise ratio (SNR), as computed by {\footnotesize PERIOD04}, was greater than 4. When such significant peaks were found the data were folded at the indicated period and two Kolmogorov-Smirnov (K-S) tests were performed. The first K-S test tested the null hypothesis that the light curve had a constant value. The second K-S test tested the null hypothesis that the light curve data had a Gaussian distribution with the mean and standard deviation of the data. For each K-S test, we report the probability, $p$, that the data represent the null hypothesis and the distance between cumulative distributions, $d$, upon which the probability is based. All K-S statistics were computed using the Numerical Recipes algorithms \citep{press1986}.

\begin{figure}
\begin{center}
\includegraphics[angle=0,scale=1]{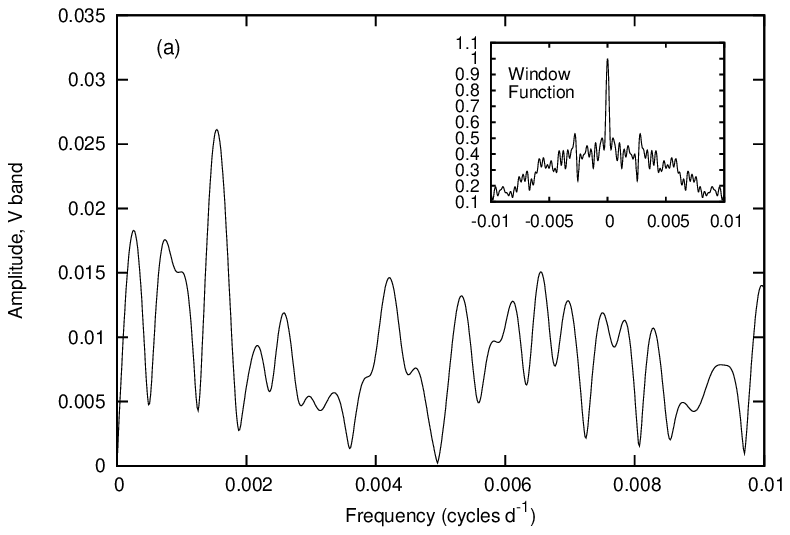}

\includegraphics[angle=0,scale=1]{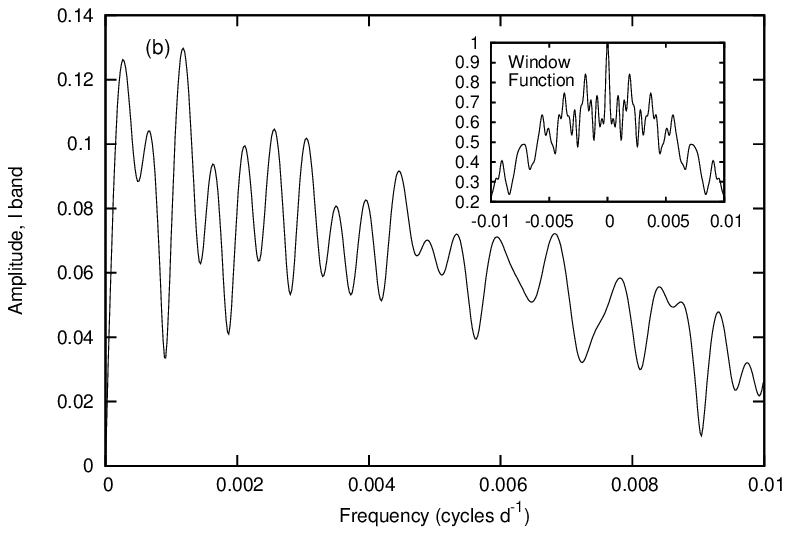}

\includegraphics[angle=0,scale=1]{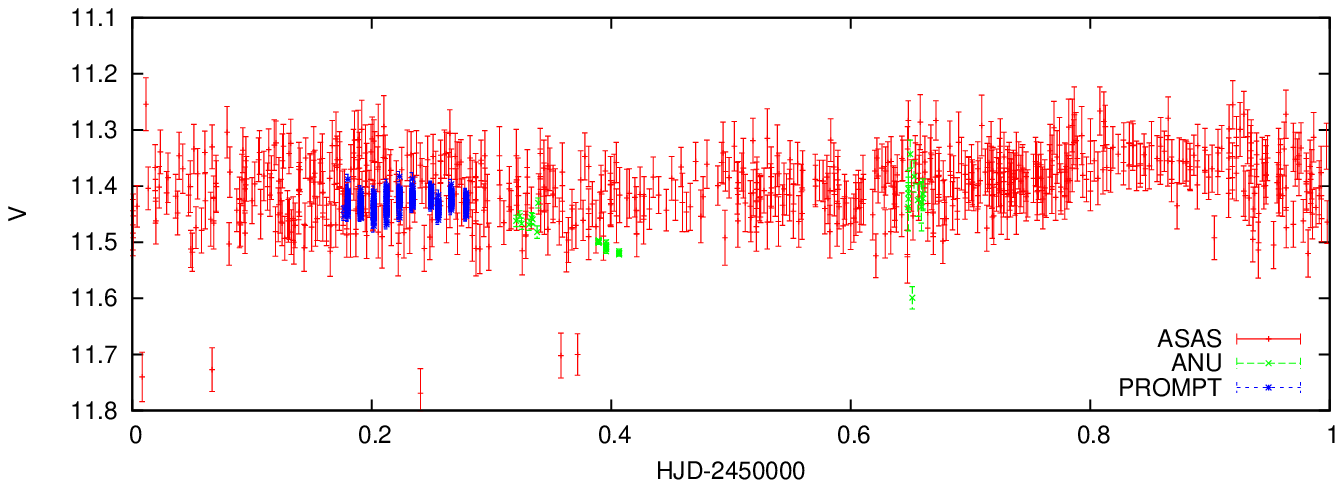}

\end{center}
\caption{(a) Periodogram for the LS 1698 $V$ band data. The peak frequency corresponds to a frequency of 0.00155 cycles d$^{-1}$, which is a period of 645~d. (b) Periodogram for the LS 1698 $I_{C}$ band data. The peak frequency corresponds to a frequency of 0.00118 cycles d$^{-1}$, which is a period of 847~d. (c) $V$ band data folded at the 645~d period. No consistent variation at the 645~d period is apparent, however note that the higher precision PROMPT data hint at a possible shorter period variation in the light curve. \label{VperLPH069fig}}
\end{figure}

\begin{figure}
\begin{center}
\includegraphics[angle=0,scale=1]{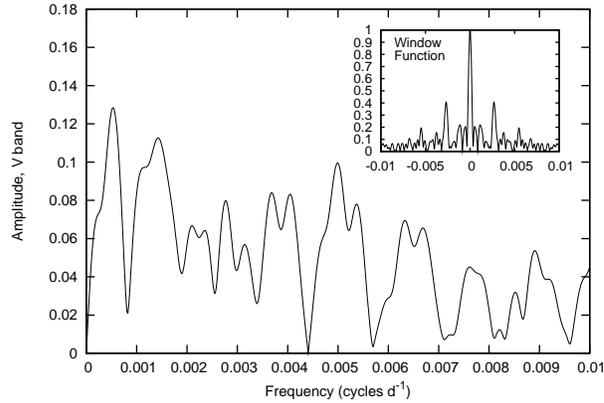}

\end{center}
\caption{Periodogram for the HD 110432 $V$ band data. The peak frequency corresponds to a frequency of 0.00054 cycles d$^{-1}$, which is a period of 1852~d.  \label{VperLPH079fig}}
\end{figure}

\begin{figure}
\begin{center}
\includegraphics[angle=0,scale=1]{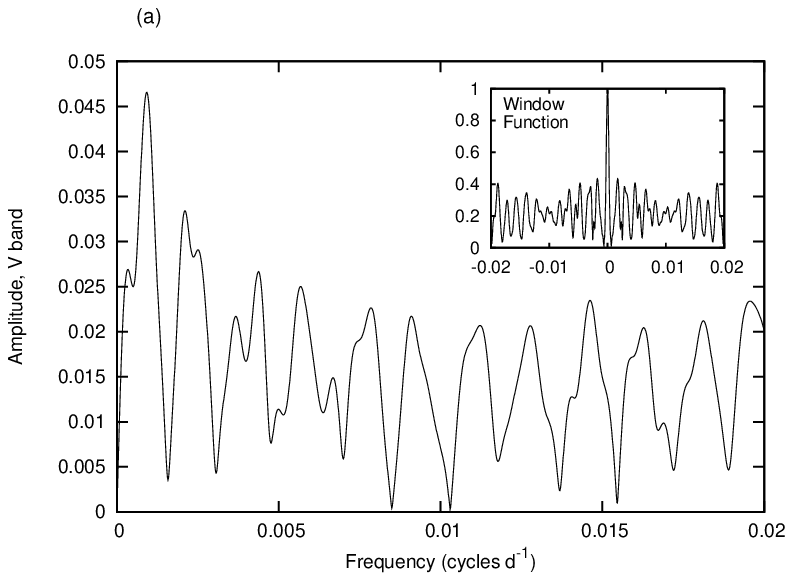}

\includegraphics[angle=0,scale=1]{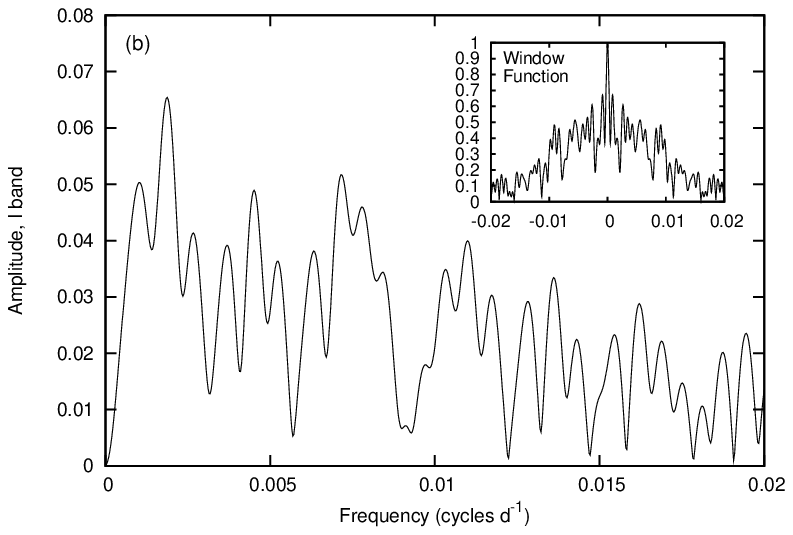}


\end{center}
\caption{(a) Periodogram for the HD 161103 $V$ band data. The peak frequency corresponds to a frequency of 0.00093 cycles d$^{-1}$, which is a period of 1075~d. (b) Periodogram for the HD 161103 $I_{C}$ band data. The peak frequency corresponds to a frequency of 0.00188 cycles d$^{-1}$, which is a period of 532~d. \label{VperLPH095fig}}
\end{figure}

For the source LS 1698, period searches were done for the entire $V$- and $I_{C}$-band datasets. It was not necessary to trim the obvious outliers (see Fig.~\ref{LPH069Vfig}) from the data prior to analysis since analysis with arbitrarily defined outliers removed did not change the results. The search results are given in Table \ref{tabpersearch} and the resulting periodograms are shown in Fig.~\ref{VperLPH069fig}. Note the discrepancy between the periods found in the $V$- and $I_{C}$-bands. A visual inspection of Fig.~\ref{VperLPH069fig} (look at the window function shape) and Fig.~\ref{LPH069Vfig} suggests that the higher frequency found for the $I_{C}$-band data is likely an artifact of the sampling gaps and that the variation of the $I_{C}$-band data roughly correlates with the variation of the $V$-band data. For example, a minimum for both light curves occurs around HJD 2453800. We therefore regarded the 645 d period as a viable period for LS 1698. Folding the data at the 645 d period did not reveal a convincing periodic variation, see
Fig.~\ref{VperLPH069fig}(c). The first K-S test did not show any evidence of variation, $d=0.002$, $p=0.99$. The second K-S test indicated that the variation of the data was significantly different from Gaussian noise, $d=0.170$, $p = 1 \times 10^{-29}$. The more densely sampled PROMPT-5 data, which show a hint of periodic variation upon visual inspection, were subject to a separate search for higher frequencies but no significant periods were found. A period search of the unweighted {\em RTXE} data for periods yielded a period of 184 d at a marginal SNR of 4.2. The period 184 d is almost exactly a half year and is likely an instrumental artifact \citep{wen2006,sarty2009}. A search with weighted data showed no significant periods.

For the source HD 110432 a formal period search was done for the $V$-band dataset. Outliers did not need to be removed from the data prior to analysis. The search results are given in Table \ref{tabpersearch} and the resulting periodograms are shown in Fig.~\ref{VperLPH079fig}. The most significant period of 1852 d essentially corresponds to the length of the data set, so continued observations for at least another 5 years are required to confirm a period of that length.

For the source HD 161103, period searches were done on the entire $V$- and $I_{C}$-band datasets, again without trimming the small number of apparent outliers from the datasets. The search results are given in Table \ref{tabpersearch} and the resulting periodograms are shown in Fig.~\ref{VperLPH095fig}. Folding the data at the 1075 d period did not reveal a convincing periodic variation. The first K-S test did not show any evidence of variation, $d=0.002$, $p=0.99$. The second K-S test indicated that the variation of the data was significantly different from Gaussian noise, $d=0.130$, $p = 1 \times 10^{-14}$.  Once again, there was a discrepancy between the periods found for the $V$- and $I_{C}$-band data. The higher frequency period found in the $I_{C}$-band data may again be an artifact of the gaps in the sampling. A visual inspection of Fig.~\ref{VperLPH095fig} reveals a possible negative correlation between the $V$- and the $I_{C}$-band light curves. We will comment on this anti-correlation in Section \ref{sec4}.

\begin{table}
\caption{Period search results. \label{tabpersearch}}
\begin{center}
\begin{tabular}{ccccc}
\hline
Source & Band & Period  & Amplitude & SNR \\
       &      &  (d)    & (mmag)    &     \\
\hline
LS 1698 & $V$  & 645 &  35       & 7 \\
        & $I_{C}$  & 847 &  155      & 18 \\
HD 110432 & $V$ & 1852 & 134       & 7 \\
HD 161103 & $V$ & 1075 & 37        & 7 \\
          & $I_{C}$ & 532 & 52        & 7 \\
\hline
\end{tabular}
\end{center}
\end{table}

\clearpage

\section{Discussion}\label{sec4}

As a part of motivating further long-term optical observations of LS 1698, HD 110432 and HD 161103, we give a brief background of what is known about these three sources.

\subsection{Background for the three systems}

\subsubsection{LS 1698 / RX J1037.5$-$5647}

The source 4U 1036$-$56 (believed to be the same as LS 1698 / RX J1037.5$-$5647 according to \citet{Motch1997}) was first observed by {\em Uhuru} \citep{Forman1978} and {\em OSO 7} \citep{Markert1979} with a mean {\em Uhuru} flux of $1.0 \times 10^{-10}$ erg cm$^{-2}$ s$^{-1}$ (2-10 keV). A 
flare appeared in November 1974, with a flux of $2.4 \times 10^{-10}$ erg cm$^{-2}$ s$^{-1}$ (2-10 keV) in
observations made by {\em Ariel V} \citep{Warwick1981}. {\em ROSAT} detected a flux about 10 times less than the {\em Uhuru} flux between 1970 and 1976 with later pointed {\em RXTE} observations by \citet{Motch1997} being about 20 times fainter than the {\em ROSAT} 1970-1976 values. The optical counterpart, LS 1698, of RX J1037.5$-$5647 was identified by \citet{Motch1997} on the basis that it was within the 95\% confidence region of the {\em ROSAT} error circle from the first {\em ROSAT} all-sky survey \citep{voges1992,motch1991} and that $L_{X}/L_{\rm bol} \geq 10^{-5}$ which was consistent with being an X-ray binary. \citet{Motch1997} further obtained blue and red spectra with the EFOSC2 instrument on the ESO-MPI 2.2-m telescope that indicate a B0 V-IIIe classification with a reddening of $E(B-V)$ = 0.75$\pm$0.25, meaning a probable distance of $\sim$5 kpc as opposed to an earlier estimate of 18 kpc. An X-ray pulse period of 860$\pm$2 s has been measured by \citet{reig1999} for LS 1698 / RX J1037.5$-$5647 in {\em RXTE} Proportional Counter Array (PCA) observations with X-ray flux levels consistent with the earlier {\em Uhuru} and {\em Ariel V} levels. Assuming a distance of 5 kpc, the X-ray flux levels measured by \citet{reig1999} translate to an X-ray luminosity of $\sim$$4.5 \times 10^{35}$ erg s$^{-1}$ (3--30 keV). The existence of X-ray pulses and a $10^{35}$ erg s$^{-1}$ flux level indicates the presence of a spinning, accreting magnetic neutron star\footnote{A neutron star radius and mass, $R = 10^{6}$ cm and $M = 1.4$ M$_{\odot}$, are implied by $L_{X} = \eta G\dot{m}M/R$ for reasonable accretion rates, $\dot{m} \sim 10^{-11}$ M$_{\odot}$ yr$^{-1}$ and $L_{x} \sim 10^{35}$ erg. Here $G$ is the universal gravitational constant and $\eta \sim 0.5$ is the efficiency with which gravitational potential energy is converted to light. Pulsations would be caused by the magnetic funnelling of accreting material onto magnetic pole hot spots on the neutron star that rotate in and out of view.} and the \citet{corbet1986} spin-orbit period relation for BeXs suggests that the orbital period for LS 1698/RX J1037.5-5647 is on the order of 200 d. Our data indicate that the orbital period is possibly 645 d. The X-ray variability appears to be around a factor of 10 with the observed flares possibly being caused by sporadic ejections of Be star disk material; with such unsteady accretion we might expect that the neutron star spin has not reached an equilibrium and therefore the system will not fit the Corbet spin-orbit relation \citep{reig1999}. The faster spin may be due to an equilibrium with the accretion of the denser clumpy material instead of with the background density of material at the the neutron star's orbit.

\subsubsection{HD 110432 / 1H 1249$-$637}

HD 110432 (= BZ Cru = HR 4830), a member of the 60 Myr old open cluster NGC 4609 behind the southern Coalsack \citep{Feinstein1979}, was identified as the optical counterpart to 1H 1249$-$637 by \citet{Tuohy1988}. Optical observations by \citet{Dachs1989} with the 61-cm University of Bochum Cassegrain telescope at La Silla, Chile supported an MK classification of B1 IIIe. Model fits to their optical spectra indicate $T_{\rm eff}=22500$ K, $\log g = 3.5$, $R_{*} = 8.6$ $R_{\cdot}$, a projected rotation velocity $v_{\rm rot} \sin i = 300$ km s$^{-1}$,  an interstellar extinction of $E^{\rm is}(B-V)=0.34$ and a circumstellar extinction of $E^{\rm cs}(B-V)=0.16$. UV observations with the {\em IUE} space telescope
by \citet{Codina1984} indicate a rotational $v_{\rm rot} \sin i$ for the primary of 360 km s$^{-1}$, $T_{\rm eff} = 25000$ K, $\log g = 3.5$, a reddening of $E(B-V)=0.40$ and spectral classification of B0.5 IIIe with an absolute magnitude of $M_{V} = -4.3$ in general agreement with the results of \citet{Dachs1989}. \citet{Codina1984} quote a distance of $430 \pm 60$ pc which is in good agreement with {\em HIPPARCOS} parallax measurements that give the distance to HD 110432 as 300$^{+60}_{-40}$ pc \citep{Chevalier1998}. The magnitude and position of the Balmer discontinuity implies $T_{\rm eff} = 22510$ K, $\log g = 3.9$ and M$_{*} = 9.6 $M$_{\odot}$ \citep{zorec2005}. The UV observations of \citet{Codina1984} further show broad photospheric absorption lines, narrow emission lines of N V, C IV and Si IV with radial velocities of $-1350$ km s$^{-1}$ and asymmetric profiles, for C IV and Si IV, that suggest an expanding envelope corresponding to a mass-loss of $3 \times 10^{-9}$ M$_{\odot}$ yr$^{-1}$. The rotational velocity of $v_{\rm rot} \sin i = 300$ km s$^{-1}$, if matched with the 1.77 d photometric period found by \citet{Barrera1991} and the radius of a B0.5 III star, 11.55 R$_{\odot}$, indicates a primary spin axis inclination of $i = 58^{\circ}$ \citep{Balona1995}. \citet{Smith2006} did not observe the 1.77 d period in their optical photometry but report a possible 130 d period.  \citet{Codina1984} also note a strong polarization of the light from HD 110432 in observations made with a photopolarimeter attached to the 1.6-m telescope of the Brazilian Astrophysical Observatory (OAB) in Brasopolis, however they attribute most of the polarization to the star's proximity to the southern Coalsack dark nebula which is between the star and the Earth \citep{Crawford1991}. 
Radial velocities were observed to vary between 27 and 70 km s$^{-1}$ by \citet{Buscombe1962}, likely based on moving emission lines; \citet{Smith2006} adopt the latest value of 6 km s$^{-1}$ from \citet{Thackeray1973}, which is confirmed by their own measurements, for their work.

Variable sub-features, wind absorptions over emission lines, have been observed the UV (C IV $\lambda$1548, Si IV $\lambda$1403 and N V $\lambda$1238) \citep{Codina1984,Smith2006}. More dramatic migrating subfeatures (msf's) have been observed in the optical HeI $\lambda$5876 and $\lambda$6678 emission lines with accelerations on the order of 100 km s$^{-1}$ hr$^{-1}$ \citep{Smith2006}. The msf's have been hypothesized to be due to strong magnetic connections between the Be star and its decretion disk \citep{Smith2006} and associated magnetic reconnection events may be the source of X-ray emission \citep{Smith1999,Robinson2002}. The optical emission lines of Fe II and He I display double-lobed profiles typical of a disk, the Be star's decretion disk, seen nearly edge-on. Using spectral synthesis techniques, \cite{Smith2006} find the temperature and density of the disk to be roughly 9800 K and $3 \times 10^{22}$ cm$^{-2}$ respectively. They also find that the projected disk covers a large 100 stellar areas out to a distance of 1 AU ($\sim$30 R$_{*}$). The inferred disk volume and mass are approximately $10^{48}$ cm$^{3}$ and $10^{-9}$ M$_{\odot}$. The strongest absorption wings in the optical and UV lines extend out to at least $\pm 1000$ km s$^{-1}$ \citep{Smith2006} which may be due to circumstellar material.

\citet{Torrejon2001} report the detection of a 14 ks X-ray pulsation period from {\em BeppoSAX} observations that they attribute to the partial occultation of a hot spot on the compact object, likely a white dwarf, as it spins. Later observations with {\em XMM-Newton} do not show a coherent 14 ks period but time-series of the 2--12keV/0.6--2keV hardness ratio do show a 14 ks period \citep{oliveria2007}. An equilibium between accretion angular momentum and spin for a white dwarf with a magnetic field of 10$^{6}$ G and a spin period of 14 ks, implies an orbital period of several hundred days \citep{Apparao1994}; our observed $V$-band period of 1852 d is longer by roughly a factor of two. However, based on the observations described in the previous two paragraphs, HD 110432 appears to have a dense and complex circumstellar environment with a large dusty disk that interacts with the primary's magnetic fields. So the density of accreting matter at the white dwarf orbit may be higher than that implicitly assumed by \citet{Apparao1994} and thus consistent with a 1852 d orbital period and an equilibrium spin period of 14 ks.

The flux observed by {\em BeppoSAX} was $L_{\rm 2-10 keV} = 3.4 \times 10^{32}$ erg s$^{-1}$ assuming the {\em HIPPARCOS} derived distance of 300 pc. Past observations of 1H 1249-637 with {\em HEAO 1} and {\em ROSAT} (it is too faint for the {\em RXTE} ASM) indicate a persistent low luminosity source consistent with a wind-accreting white dwarf\footnote{An accretion rate of $\dot{m} = 4 \times 10^{-11}$ M$_{\odot}$ yr$^{-1}$ is implied by $L_{x} = 3.4 \times 10^{32}$ erg s$^{-1}$, $\eta = 0.5$ and a white dwarf with $R = 10^{9}$ cm and $M = 1.0$ M$_{\odot}$.} \citep{Torrejon2001,Waters1989}. 1H 1249-637 shows an emission feature at 6.8 keV with an equivalent width of 600 to 700 eV which is too large to be associated with an accreting neutron star but is similar to that found in some cataclysmic variables \citep{Torrejon2001}, which contain white dwarfs. Observations with {\em XMM-Newton} show the 6.8 keV feature to be an Fe K$\alpha$ complex of three lines (at 6.4, 6.7 and 6.97 keV) plus there is a suspected Fe XXVI Ly$\beta$ emission at 8.2 keV \citep{oliveria2007}. \citet{Torrejon2001} note that $E(B-V)=0.40$ implies an absorption of $N_{H} \sim 0.3 \times 10^{22}$ cm$^{-2}$ but $N_{H} \sim (1.1$--$1.4) \times 10^{22}$ cm$^{-2}$ is deduced from the X-ray spectral model fits which imples a higher density of circumstellar material around the X-ray source than around the optical source. The existence of denser material around an accreting compact object may, again, be responsible for a relatively faster spin period of 14.5 ks at a relatively slower 1852 d orbital period. \citet{oliveria2007} find that a three thermal plasma model best fits their {\em XMM-Newton} X-ray spectra; the three plasmas have $kT$ temperatures of [0.2--0.7 : 3--6 : 16--37] in a [few\% : 15--25\% : 70--80\%] distribution. The soft component is consistent with a normal B star shocked wind emission \citep{oliveria2006}. The hard component qualifies it, along with the low X-ray luminosity and the Fe K$\alpha$ complex, as a $\gamma$ Cas analogue. The three components of the X-ray spectra again attest to the complex circumstellar environment of HD 110432. No coherent 14.5 ks variations were found, but the {\em XMM-Newton} 0.6--2 and 2-12 keV light curves examined by \citet{oliveria2007} showed `flaring' activity characterized by a $\sim$$f^{-0.75}$ power spectra. The brightness of the flares was not correlated with their X-ray hardness. \

\subsubsection{HD 161103 / RX J1744.7$-$2713}

HD 161103 was identified as the optical counterpart of the BeX candidate RX J1744.7$-$2713 by \citet{Motch1997}. The identification was made on the basis of optical spectroscopic observations with the ESO-MPI 2.2-m telescope, which showed Balmer emission lines, and pointed observations with {\em ROSAT} to follow up observations from the {\em ROSAT} Galactic Plane Survey. HD 161103 is a B0.5 V-IIIe star with a hard ($\gamma$ Cas-like) X-ray emission with an unabsorbed luminosity of $\sim$10$^{32}$ erg s$^{-1}$ and $L_{\rm X}/L_{\rm bol}$ of about $4 \times 10^{-6}$ \citep{Motch1997,Steele1999,oliveria2006}. Similar to HD 110432, its X-ray spectra as observed with {\em XMM-Newton} \citep{oliveria2006} could be equally well-fit with a optically thin hot plasma, with $T \sim$10$^{8}$ K and solar abundances ({\footnotesize XSPEC} mekal model), or by a power law with photon index $\Gamma \sim$1.5--1.8. The X-ray spectrum also shows an Fe K$\alpha$ complex with a neutral fluorescent line at 6.4 keV, an He-like Fe XXV emission at 6.7 keV and an H-like Fe XXVI emission at 6.97 keV. Such Fe K$\alpha$ complexes are not seen in BeXs where there is direct accretion onto the surface of a neutron star and the 6.4 keV flouorescence indicates a reprocessing of radiation in cooler circumstellar matter. The H- and He-like Fe lines indicate a hot plasma compatible with the overall mekal fit \citep{oliveria2006}. A two component model with hot and cold components also fits the spectrum well with the hot component being less absorbed. \citet{oliveria2006} observed a coherent sinusoidal oscillation with a period of 3245$\pm$350 s over most of a 17.6 ks observation in the 0.5--12 keV band that may be related to the spin of a compact object. If so then the spin-orbit relation of \citet{Apparao1994} for Be/white dwarf systems would imply an orbital period of a few hundred days and, again, we would explain the longer period of 1075 d implied by our data as being due to a denser circumstellar environment than implicitly assumed by Apparao. A white dwarf magnetic field less than the $10^{6}$ G assumed by Apparao would also lead to a longer orbital period for a given white dwarf spin rate. In assessing any proposed spin-orbit relations for both HD 110432 and HD 161103 it should be remembered that the X-ray pulsations were observed only once for each object; both spin periods still need further observation to confirm their existence.

The B0.5 classification along with a luminosity class of V or III imply a distance to HD 161103 of 1.1 or 2.0 kpc respectively \citep{oliveria2006,Wegner1994,Humphreys1984}.
The optical spectra of HD 161103 show a large equivalent width H$\alpha$ line, -31 to -34 \AA, with weak V/R asymmetry along with the Paschen line series in emission in the IR spectra that indicates the presence of a large stable decretion disk \citep{oliveria2006}. The large disk again points towards larger possible accretion onto a white dwarf than implied by the relation of \citet{Apparao1994}. The O I$\lambda$8446 line, excited by Ly$\beta$ photons \citep{Andrillat1988}, is also seen in emission. An EW(H$\alpha$)-$P_{\rm orb}$ relation discovered by \citet{Reig1997} implies $P_{\rm orb} > 100$ d. A number of Fe emission lines from the circumstellar environment are visible while the He I $\lambda$4388 photospheric line varies by less than 13 km s$^{-1}$ \citep{oliveria2006}. \citet{Steele1999} find $v_{\rm rot} \sin i = 244 \pm 33$ km s$^{-1}$ from the He I lines. \citet{oliveria2006} note that the strength of the N II $\lambda$3995, 4044 lines indicate that HD 161103 is moderately Nitrogen-rich, a quality seen in other HMXB primaries that could be the result of mass transfer in binary evolution -- a fact that argues for the binary nature of this $\gamma$ Cas analogue.

\subsection{Further searches and possible causes for orbitally induced optical variations}

Based on previous observations of the these, and other HMXBs, all three of the HMXBs we observed are expected to have long orbital periods, if they are, in fact, binary systems. Although we are unable to confirm any periods in the optical data we have so far, further long term observations, at a precision of 0.01 mag or better (e.g. at the the precision of our PROMPT data), may yet reveal periodic variation. If such variation is driven by orbital motion, it would raise the question of what physical mechanism is behind the the light curve variation. Ordinarily, two mechanisms are responsible for orbital period light curve variation: eclipses and ellipsoidal variation. Variation due to eclipses would be one or two localized dips in the brightness of the star, due to the primary and secondary eclipses, and not the smooth variation seen in our data. Ellipsoidal variation, caused by the primary being distorted into an ellipsoid-like shape by gravitational tidal forces, generally requires a short orbital period or a highly eccentric one with a small periastron distance. The orbital periods for the three sources studied here here are likely not short and a close periastron would result in regular X-ray outbursts which are not seen. Optical modulation can occur in synchrony with X-ray modulation with both bands brightening, possibly with the optical brightening occurring first, in response to increased mass flow through an accretion disk as the compact object passes perihelion \citep{Schmidtke2006,Kriss1983}. Where we have X-ray data, for LS1698, there is no correlation of the X-ray output with the optical light curve, ruling out simple accretion disk models in that case. The X-ray luminosities of the other two sources HD 110432 and HD 161103 are too low to produce a long-term record in {\em RXTE} data. So it may yet be true that the low luminosity X-ray light curves of HD 110432 and HD 161103 correlate with their optical light curve.

In the absence of the modulation of the optical light curve by direct accretion onto the compact object, a mechanism other than eclipses and ellipsoidal variation would have to be responsible for the variation seen in our data for LS 1698, HD 110432 and HD 161103. For example, decretion disk activity or the interaction of the decretion disk with a white dwarf as causes for ``type 1'' Be star activity could be responsible \citep{mennickent2001}. However such type 1 activity generally takes the form of an optical outburst in contrast with the slower variation seen in our data. We therefore investigated two, speculative, geometric models for orbitally induced optical variation here without delving into the detailed physical mechanisms behind the models. With the geometry narrowed down by at fit of these models to a future data set we may then speculate about the required physical mechanisms. The two models we investigated are the `emitting shock model' and the `hot surface model'.

In the emitting shock model it was assumed that a shock preceding the neutron star (or a possible white dwarf for the cases of HD 110432 and HD 161103) emits light by some unspecified physical process. Colliding winds from the compact star and the primary star may give rise to such a shock or the shock might be at the boundary of the compact object's magnetosphere. Generally such shocks would be expected to emit high energy $\gamma$ or X radiation (see e.g. models for the HMXBs LS 5039 and LSI +61$^{\circ}$303 \citep{Cerutti2008}) but here we allow for the possibility of optical band emission. For simplicity, the emitting shock was modelled as a half-lit surface of a sphere aligned with the orbital velocity vector of the neutron star. The half-lit surface would thus present a well-defined phase to the observer with the maximum of the light curve corresponding to the full phase, or at least maximum visible phase, that occurs when the neutron star is approaching the Earth. 

In the hot surface model it was assumed that the half of the primary star facing the neutron star is brighter than the other half. In this model the light curve maximum occurs when the neutron star is between the primary star and the Earth when the hot surface presents a full phase, or a maximum visible phase, to the Earth. To re-emphasize, the model is purely geometric; there is no modelling of brightness as a function of the inverse square distance between the component stars for example. A similar hot surface model, with an X-ray shadowing disk, has been proposed for the X-ray binary Her X-1 \citep{Gerend1976,Howarth1983}.

Both models produced light curves that visually match the variation seen in the unfolded light curves of Figs.~\ref{LPH069Vfig} to \ref{LPH095Vfig}. However, until convincing values for the orbital periods are found, it makes little sense to formally fit the two geometric models to the data.

The variation in the $V$-band magnitude of the lightcurves implies a variation in the energy outputs of the systems that we may estimate given the distances to the sources. Specifically we may estimate the luminosity of the system $L_{*}$ at maximum and minimum brightness via
\begin{equation}
L_{*} = L_{\odot} \left(  \frac{d_{*}}{d_{\odot}} \right)^{2} 10^{(V_{\odot}-V_{*}+A_{v})/2.5}
\end{equation}
and take the difference to estimate the luminosity change. Here $L_{\odot}$, $d_{\odot}$ and $V_{\odot}$ are the solar luminosity, distance and apparent $V$-band magnitudes respectively; $L_{*}$, $d_{*}$ and $V_{*}$ are the same parameters for the system and $A_{v} = 3.2 E(B-V)$ is the interstellar absorption. Using quantities found by others, as reported in the background sections above, we find a change in energy output of $1.3 \times 10^{36}$ erg s$^{-1}$ for LS 1698, $9.6 \times 10^{35}$ erg s$^{-1}$ for HD 110432 and $1.1 \times 10^{35}$ erg s$^{-1}$ for HD 161103. These values very nearly match the observed X-ray luminosity for LS 1698 and are brighter than the X-ray luminosity of the other two systems. This suggests that the X-ray output of the system is reprocessed to optical wavelengths when the sources are optically bright, if an argument can be made that nearly all of the X-ray output is intercepted by the reprocessing medium. In the case of HD 110432 and HD 161103, additional, or alternate, sources of energy are also implied. This difference in energy source between LS 1698 and the pair HD 110432 and HD 161103 further supports the notion that the character of the systems are different, recalling that LS 1698 likely contains a neutron star and the other two systems likely contain a white dwarfs.

In order for the X-ray energy to be reprocessed to optical wavelengths, an optically thick reprocessing media is necessary. An emitting shock is likely to be optically thin and, if that is the case, it would rule out such a shock as a reprocessing source of the systems' variable optical output. The hot surface, if on the primary, would have to be on the cooler surface of the star. Another possibility for a reprocessing site could be protrusions of a cold disk. It may be possible, for example, that the geometry of an illuminated cold disk edge is similar to that of our hot surface model.

Simpler explanations for the observed optical variation could be related to the expected dusty circumstellar environments of BeXs. The primary Be stars likely have equatorial dusty decretion disks, as indicated in their optical spectroscopy (see the backgrounds sections above). The interaction of the compact object with that dusty circumstellar environment may produce the observed optical variations as denser regions near the neutron star (or white dwarf) obscure the light from the primary. This obscuring dust model has some support from the $V$- and $I_{C}$-band light curves of HD 161103 (see Fig.~\ref{LPH095Vfig}); as the $V$-band light curve gets brighter, the $I_{C}$-band light curve appears to dim. An obvious explanation is that as the dust clears, the source gets brighter in the $V$-band and as the dust obscures the primary we see that dust as increased $I_{C}$-band brightness.

Clearly, more optical band data, of a higher precision than those of the ASAS data analysed here, are required before any of the models outlined here can be considered further. 

\section{Conclusion}\label{sec5}

Long term variation is evident in $V$-band light curve data for three HMXBs. Using distances determined by others, the variation in optical output from the three HMXB systems implies a variation in luminosity of $\sim$10$^{35}$ erg s$^{-1}$ in all three cases. This variation in luminosity matches the X-ray luminosity of LS 1698 which is suspected of containing a neutron star. The X-ray luminosity reported by others for HD 110432 and HD 161103, which are suspected of containing accreting white dwarfs, is smaller at $\sim$10$^{32}$ erg s$^{-1}$. The light curve variations are roughly consistent with a hot surface source on the hemisphere of the primary facing a compact accreting object or with a hot shock-like structure preceding the compact object in its orbit. The variation may also, or instead, be due to obscuration by circumstellar dust and gas as denser regions follow the orbit of the compact object. The hypothesis of variation due to dust is supported by the observation of apparently anti-correlated $V$- and $I_{C}$-band light curves in the case of HD 161103. A $V$-- $I_{C}$-band anti-correlation is not apparent for the other two sources. The lack of periodicity in the 2-10 keV X-ray light curve for LS 1698 rules out a simple accretion disk brightening model for that source.

\section*{Acknowledgments}

The Digitized Sky Surveys
used to produce Figs.~\ref{stdstars69} and \ref{stdstars95} were produced at the Space Telescope
Science Institute under U.S. Government grant NAG W-2166. The
images of these surveys are based on photographic data obtained
using the Oschin Schmidt Telescope on Palomar Mountain and the
UK Schmidt Telescope. The plates were processed into the present
compressed digital form with the permission of these institutions. The Beowulf cluster used to investigate the geometrical models was constructed from recycled computers donated by the Arts and Science Computer Laboratory of the University of Saskatchewan. Thanks to Arne Henden of the AAVSO for the SRO data used to define standard stars in the field of HD 161103. Thanks to the ANU technical support staff and the wonderful cooks at the ANU lodge for their support and hospitality during the observing runs with the 24- and 40-in telescopes at Siding Spring Observatory.

\label{lastpage}

\end{document}